\documentstyle[multicol,aps,epsf]{revtex}
\def\twobeone{\end{multicols}
\vskip.6pc
\noindent
\vrule width3.375in height.2pt depth.2pt \vrule depth0em height1em\hfill 
\vskip.6pc}

\def\onebetwo{
\vskip.6pc
\indent
 \hfill\vrule depth1em height0pt \vrule width3.375in height.2pt depth.2pt
\vskip.6pc
\begin{multicols}{2}\noindent}

\def\tup{\begin{array}{c}\uparrow\\[-1mm] -\end{array}}
\def\tdown{\begin{array}{c}\downarrow\\[-1mm]-\end{array}}
\def\bup{\begin{array}{c}-\\[-1mm]\uparrow\end{array}}
\def\bdown{\begin{array}{c}-\\[-1mm]\downarrow\end{array}}

\def\braket#1{\left|#1\right>}

\begin{document}

\draft

\title
{The Ground State and Excitations of Spin Chain with Orbital Degeneracy}

\author{
You-Quan Li$^{1,2}$, Michael Ma$^1$, Da-Ning
Shi$^{1,3}$, and Fu-Chun Zhang$^1$} 
\address
{
${}^{1}$ Department of Physics, University of Cincinnati, Cincinnati OH 45221\\
${}^{2}$ Zhejiang Institute of Modern Physics,
Zhejiang University, Hangzhou 310027, China\\
${}^{3}$ College of Science, Nanjing Univ. of Aeronautics and Astronautics,
Nanjing, China 
}

\date{ Received Feb. 16, 1999}

\maketitle

\begin{abstract}
The one dimensional Heisenberg model in the presence of orbital degeneracy
is studied at the $SU(4)$ symmetric point by means of Bethe ansatz.  
Following Sutherland's previous work on an equivalent model, we  discuss
the ground state and the low-lying excitations more extensively in
connection to the spin systems with orbital degeneracy.  We show explicitly
that the ground state is a $SU(4)$ singlet. 
We study the degeneracies of the elementary
excitations and the spectra of the generalized magnons consisting
of these excitations.  
We also discuss the complex 2-strings in the context of
the Bethe ansatz solutions. 
\end{abstract}

\begin{multicols}{2}

\section{Introduction}

There has been much interest recently in spin 
Hamiltonians with orbital degeneracy. 
The orbital degree of freedom may be
relevant to many  transitional metal 
oxides~\cite{McWh,Cast,Word,Kuge,Bao,Rice,Pen,Yama,Ishi,Shii,Fein}.  
Examples of such systems in 1-dimension include
quasi-one-dimensional
tetrahis-dimethylamino-ethylene
(TDAE)-C$_{60}$~\cite{Arovas},
and artificial quantum dot arrays~\cite{Marston}.
Recently, we discussed these 
systems ~\cite{UC98} within the framework 
of a $SU(4)$ theory. 
A quantum disordered ground state in 2-dimension
was proposed to
be relevant to  the experimentally
observed unusual magnetic properties in $LiNiO_2$.   
There have also been numerical studies of
the 1-dimensional
models for these systems~\cite{Ueda,Mila}.   

In the present paper, we use the Bethe ansatz method to study the
$SU(4)$ symmetric Heisenberg spin chain with two-fold orbital degeneracy.
This model is equivalent to the model studied by
Pokrovskii and Uimin \cite{Uimin}, and to one of a class of models that has
been solved by Sutherland\cite{Sutherland75}.  Expanding on Sutherland's
work, we study the ground state and low-lying excitations more extensively by
considering holes and 2-strings in the thermodynamics limit, and in
connection to the spin systems with orbital degeneracy. We show explicitly
that the ground state is a $SU(4)$ singlet, consistent with the generalized
Lieb-Mattis theorem of Affleck and Lieb \cite{AffleckLieb}.  We discuss the
degeneracies of the elementary excitations and the spectra of the generalized
magnons resulting from such such excitations.  We also discuss the
complex 2-strings in the context of
the Bethe ansatz solutions. The paper is
organized as follows. In section II,
we introduce the model and discuss its
symmetry. In section III, we present
the Bethe ansatz solution following
Sutherland's approach. We discuss
the ground state in section IV, and the
elementary excitations and the generalized magnon modes in section V. A brief
summary is given in section VI.
\section{Symmetry consideration}

We consider the spin chain of N sites with two-fold
orbital-degeneracy and with periodic boundary condition  \cite{Rice}
\begin{equation}
{\cal H}=\sum_{j=1}^N J\left[
      (2 \vec {T}_j\cdot \vec{T}_{j+1} + \frac{1}{2})
      (2 \vec S_j\cdot \vec S_{j+1} + \frac{1}{2})
       - 1\right],
\label{eq:Hamiltonian}\end{equation}
where $\vec S_j$ and $\vec T_j$ 
are the spin and orbital operators respectively 
on the $j$-th site, and are 
each generators of a $SU(2)$ Lie algebra.
Clearly, in addition to the permutation symmetry
($H$ is invariant under the transformation 
$j \rightarrow j+1$ for all $j's$),
the system has an explicit  $SU(2)\otimes SU(2)$ global
symmetry 
and a bi-symmetry $\vec S_j \leftrightarrow \vec T_j$.
Actually the Hamiltonian (\ref{eq:Hamiltonian}) is invariant
under a global $SU(4)$ transformation, which is generated by 
$S^\alpha = \sum_j S_j^\alpha$, \,\, $T^\alpha = \sum_j T_j^\alpha$,
and $Y^{\alpha \beta} = \sum_j T_j^\alpha \otimes S_j^\beta$ 
$ (\alpha, \beta = 1, 2, 3)$. These operators satisfy 
the following commutation
relations   \begin{eqnarray}
[S^\alpha, S^\beta ] &=& i\epsilon^{\alpha\beta\gamma}S^\gamma,
   \,\,\,\, [S^\alpha, T^\beta ] =0, \nonumber\\[1mm]  
[T^\alpha, T^\beta ] &=& i\epsilon^{\alpha\beta\gamma}T^\gamma,\,\,
[T^\alpha, Y^{\beta\delta}] = i\epsilon^{\alpha\beta\gamma}
   Y^{\gamma\delta},             \nonumber\\[1mm]
[S^\alpha, Y^{\delta\beta}] &=& i\epsilon^{\alpha\beta\gamma}
   Y^{\delta\gamma},              \nonumber\\[1mm]
[Y^{\alpha\mu}, Y^{\beta\nu}] &=& i\epsilon^{\alpha\beta\gamma}
   \delta^{\mu\nu}S^{\gamma} + i\epsilon^{\mu\nu\rho}
    \delta^{\alpha\beta}T^\rho.
\label{eq:commutators}
\end{eqnarray}
Such a symmetry was noticed by Wigner in the study
of nuclei long time ago\cite{Wigner}. 
Since the group $SU(4)$ is of rank three, 
there are three conserved quantum
numbers in general. 
It is useful to write the above commutation relations in terms
of Chevalley basis, i.e.,
three generators in the Cartan sub-algebra of $SU(4)$
(precisely, the $A_3$ Lie algebra) 
$H_n =\sum_{j=1}^{N}H_n(j)$ $(n=1,\, 2,\, 3)$
which can be diagonalised simultaneously, and the other 
twelve generators $E_\alpha = \sum_{j=1}^{N}E_\alpha(j)$ 
($\alpha$ denotes root vectors).
The local generators $H_n(j)$ and $E_{\alpha}(j)$ are related to  
the spin and orbital operators by,
\begin{eqnarray}
H_1(j) &=& (1+2T^z_j)S^z_j , \,\,\, 
E_{\alpha_1}(j) = (\displaystyle\frac{1}{2}+T^z_j)S^+_j ,\nonumber\\
H_2(j) &=& (T^z_j -S^z_j), \,\,\,\,\,\,\,\,\,\,     
E_{\alpha_2}(j) = T^+_j S^-_j,   \nonumber\\
H_3(j) &=& (1-2T^z_j)S^z_j,  \,\,\,    
E_{\alpha_3}(j) = (\displaystyle\frac{1}{2}-T^z_j)S^+_j, \nonumber\\ 
\, &\,& 
  E_{\alpha_1+\alpha_2}(j) 
   =T^+_j(\displaystyle\frac{1}{2}+S^z_j),  \nonumber\\
\, &\,& 
  E_{\alpha_2+\alpha_3}(j) 
   =T^+_j(\displaystyle\frac{1}{2}-S^z_j), \nonumber\\ 
\, &\,& E_{\alpha_1+\alpha_2 +\alpha_3}(j)=T^+_j S^+_j, 
\label{eq:chevalley}
\end{eqnarray}
where $\alpha_1$, $\alpha_2$ and $\alpha_3$ denote the 
simple roots of $A_3$ Lie algebra\cite{Gilmore}. 
The generators corresponding to the negative roots 
are given by
$E_{-\alpha}(j) = E_\alpha^\dagger(j)$.
One can verify that these operators satisfy the standard
commutation relations of $A_3$ Lie algebra, and 
Eq.(\ref{eq:Hamiltonian}) can be rewritten as,
$${\cal H} = \sum_j[h(j,j+1)-3/4],$$
$$h(j, j+1)=\sum_{m,n}g^{m n}H_m(j) H_n(j+1)
    +\sum_{\alpha\in\Delta}E_\alpha (j) E_{-\alpha}(j+1),$$
where $\Delta$ denotes the set of roots of the Lie algebra.
We denote the spin components by up ($\uparrow$) and down ($\downarrow$),
and the orbital components by top and bottom. Then the four possible states 
on each site are
\begin{eqnarray} 
|1>&:=&|\tup >=|1/2, 1/2>, \nonumber\\ 
|2>&:=&|\tdown >=|-1/2, 1/2>, \nonumber\\ 
|3>&:=&|\bup >=|1/2, -1/2>, \nonumber\\ 
|4>&:=&|\bdown >=|-1/2, -1/2>. \nonumber
\end{eqnarray}
The local lowering/raising operators
$E_{\pm\alpha_n}(j)$ 
 relate those four states on the $j$th site as follows,
\begin{eqnarray}
E_{-\alpha_n}(j) |n>_j  & \rightarrow & |n+1>_j, \nonumber\\
E_{\alpha_n}(j) |n+1>_j & \rightarrow & |n>_j, \,\, 
n= 1, 2, 3.
\end{eqnarray} 
In terms of those operators,
a general state can be written as 
\twobeone
\begin{equation}
|\psi>=\sum_{\{x_i\}\{y_j\}\{z_k\}}
     \psi (x; y; z)
       \prod_{k=1}^{M''}E_{-\alpha_3}(z_k)
        \prod_{j=1}^{M'}E_{-\alpha_2}(y_j)
          \prod_{i=1}^{M}E_{-\alpha_1}(x_i)
           \braket{\tup\tup\cdots\tup},
\label{eq:generalstate}
\end{equation}
\onebetwo
where $x:=(x_1, x_2, \cdots, x_M)$, 
$y:=(y_1, y_2,\cdots, y_{M'})$,
$z:=(z_1, z_2,\cdots, z_{M''})$;
$1\leq x_1 < x_2 < \cdots < x_M \leq N$;
$x_1 \leq y_1 < y_2 < \cdots < y_{M'} \leq x_M $, 
$y_1 \leq z_1 < z_2 < \cdots < z_{M''} \leq y_{M'}$; 
and $\{z_k\}\subset\{y_j\}\subset\{x_i\}$.
We may define the weights as the eigenvalues of the global operator
$H_n$, indicated by $(H_1,H_2, H_3)$.   The eigenvalues of the local
operators $H_1(j), H_2(j), H_3(j)$ acting on the four local states
$|\tup>_j , |\tdown>_j , |\bup>_j $ and $|\bdown>_j$ are
$(1, 0, 0),$ $(-1, 1, 0),$ $(0, -1, 1)$ and $(0, 0, -1)$ respectively.
We shall focus on the 
state with the highest weight.  The other states in
the same irreducible representation 
can then be obtained by using the
corresponding lowering  
operators  $E_{-\alpha_n}$. 
In the present model, the irreducible representation of the
$SU(4)$ group of a N-site system is labeled by
\begin{equation}
(N+M'-2M, M+M''-2M',M'-2M'').
\label{eq:weight}
\end{equation}

\section{The Bethe ansatz solution}

The permutation and the $SU(4)$ symmetries in the Hamiltonian
enable us to seek the eigenstate of both the
cyclic permutation operator
and the generators of the Cartan subalgebra of 
$A_3$. The invariance of the cyclic permutation 
imposes a periodic boundary condition on the 
wave function $\psi(x,y,z)$.
The present model is solvable~\cite{Sutherland75},
and the Bethe ansatz equations for the spectra are
\begin{eqnarray}
\left( \frac{\lambda_j+i/2}{\lambda_j -i/2}\right)^N
   &=& -\prod_{l=1}^{M}
        \frac{\lambda_j-\lambda_l + i}
             {\lambda_j-\lambda_l - i}
          \prod_{\beta=1}^{M'} 
           \frac{\mu_{\beta} -\lambda_j + i/2}
                {\mu_{\beta} -\lambda_j - i/2},
        \nonumber\\
\prod_{l=1}^{M}
  \frac{\mu_\gamma - \lambda_l + i/2}
       {\mu_\gamma - \lambda_l - i/2}
   &=& -\prod_{\beta=1}^{M'}
       \frac{\mu_\gamma - \mu_\beta + i}
            {\mu_\gamma - \mu_\beta - i}
         \prod_{b=1}^{M''}
           \frac{\nu_b - \mu_\gamma + i/2}
                {\nu_b - \mu_\gamma - i/2},
         \nonumber\\       
\prod_{\beta=1}^{M'}
  \frac{\nu_c -\mu_\beta + i/2}
       {\nu_c -\mu_\beta - i/2}
    &=& -\prod_{b=1}^{M''}
         \frac{\nu_c - \nu_b + i}
              {\nu_c - \nu_b - i},
\label{eq:BAE}\end{eqnarray}
where $j, l=1, 2,\cdots, M$; $\beta, \gamma=1, 2,\cdots, M'$
and $b, c= 1, 2,\cdots, M''$. 
These are secular equations for the spectra of 
$SU(4)$ rapidities $ \lambda, \, \mu$ and $\nu$.
The energy spectrum is given by 
\begin{equation}
E=-\sum_{l=1}^M\frac{J}{(1/2)^2+\lambda_l^2}.
\end{equation}
The momentum defined by the translation of the system along
the chain is given by 
\begin{eqnarray}
P &=&\frac{1}{i}\ln\prod_{l=1}^M\frac{\lambda_l+i/2}{\lambda_l-i/2}
  \nonumber\\
\,&=&  \sum_{l=1}^M [\pi -2\tan^{-1}(2\lambda_l)].
\label{eq:momentumf}
\end{eqnarray}
Note that $P$ in eq. (\ref{eq:momentumf}) is determined up to mod ($2\pi$),
and the inverse trigonometric 
function is defined in the
main branch.
We have included explicitly the $\pi$ term in
eq.(\ref{eq:momentumf}),
which is usually neglected in the study
of pure spin Heisenberg
models\cite{Faddeev}.
In the $SU(4)$ model,
there are three types of
elementary excitations
as we will discuss below,
and it is convenient to include
the $\pi$ term in $P$ to study
the magnon types of composite
excitations.
We define the momentum of the elementary 
excitations as the momentum relative to
the ground state\cite{Faddeev}.
By taking the logarithm of (\ref{eq:BAE}), 
a set of coupled transcendental equations
are obtained,
\twobeone
\begin{eqnarray}
\Theta_{1/2}(\lambda_j) 
  -\frac{1}{N}\sum_{l=1}^{M}\Theta_1(\lambda_j-\lambda_l)
   -\frac{1}{N}\sum_{\beta}^{M'}
     \Theta_{1/2}(\mu_{\beta}-\lambda_j)
  =\frac{2\pi}{N}I_j,
   \nonumber\\
\sum_{l=1}^{M}\Theta_{1/2}(\mu_{\gamma}-\lambda_l)
   -\sum_{\beta=1}^{M'}\Theta_1(\mu_{\gamma}-\mu_{\beta})
    -\sum_{b=1}^{M''}\Theta_{1/2}(\nu_b -\mu_\gamma)
  =2\pi J_\gamma,
    \nonumber\\
\sum_{\beta=1}^{M'}\Theta_{1/2}(\nu_c -\mu_\beta)
   -\sum_{b=1}^{M''}\Theta_1 (\nu_c - \nu_b)
  =2\pi K_c, 
\label{eq:logBAE}\end{eqnarray}
\onebetwo
where $\Theta_{\rho}(x):=2\tan^{-1}(x/\rho)$.
The quantum number $I_j$ is an integer or half-integer 
depending on whether $N-M-M'$ is odd or even, and so is 
$J_\gamma$ (or $K_c$) depending on whether $M-M'-M''$
(or $M'-M''$) is odd or even. These properties arise
from the logarithm function.

Replacing $\lambda_j$, $\mu_\gamma$ and $\nu_c$ in eq. (\ref{eq:logBAE})
by continuous variables $\lambda$, $\mu$ and $\nu$ 
but keeping the summation still over the solution set of these roots
$\{\lambda_l, \mu_\beta, \nu_b \}$, we can consider the quantum
numbers $I_j$, $J_\gamma$ and $K_c$ as functions 
$I(\lambda)$, $J(\mu)$ and $K(\nu)$ given by 
eq. (\ref{eq:logBAE}). Take $I(\lambda)$ as an example. When
$I(\lambda)$  passes through one of the quantum numbers
$I_j$, the corresponding $\lambda$ is equal to
one of the roots $\lambda_j$. Similarly for $J(\mu)$ or $K(\nu)$. However,
there may exist some integers or half-integers for which the
corresponding $\lambda$ ($\mu$ or $\nu$) is not in the set of roots.
We shall name such a state as a ``hole''. In the thermodynamics limit 
$N \rightarrow \infty$, we may introduce the density of roots and
the density of holes (indicated by a subscript $h$), 

\begin{eqnarray}
\sigma(\lambda)+\sigma_h(\lambda) = (1/N)dI(\lambda)/d\lambda,
     \nonumber\\
\omega(\mu)+\omega_h(\mu) =(1/N)dJ(\mu)/d\mu, 
     \nonumber\\ 
\tau(\nu)+\tau_h(\nu) =(1/N)dK(\nu)/d\nu.
     \nonumber
\end{eqnarray}
By replacing the summations by integrals,

\begin{eqnarray}
\lim_{N\rightarrow\infty}\frac{1}{N}\sum_{l=1}^M f(\lambda_l)
=\int^B_{-B}d\lambda\sigma(\lambda)f(\lambda),
            \nonumber\\
\lim_{N\rightarrow\infty}\frac{1}{N}\sum_{\beta=1}^{M'} f(\mu_\beta)
=\int^{B'}_{-B'}d\mu\,\omega(\mu)f(\mu),
            \nonumber\\  
\lim_{N\rightarrow\infty}\frac{1}{N}\sum_{b=1}^{M''} f(\nu_b)
=\int^{B''}_{-B''}d\nu\tau(\nu)f(\nu),
\nonumber
\end{eqnarray}
eq.(9) become
the coupled integral
equations,
\twobeone
\begin{eqnarray}
\sigma(\lambda) + \sigma_h(\lambda) &=&
   K_{1/2}(\lambda) 
   - \int_{-B}^{B}d\lambda'K_1(\lambda-\lambda') \sigma(\lambda')
   + \int_{-B'}^{B'}d\mu'K_{1/2}(\lambda-\mu')\omega(\mu'),
\nonumber\\
\omega(\mu) + \omega_h(\mu) &=&
  \int_{-B}^{B}d\lambda' K_{1/2}(\mu-\lambda')\sigma(\lambda')
  -\int_{-B'}^{B'}d\mu' K_1(\mu-\mu')\omega(\mu')
   +\int_{-B''}^{B''}d\nu' K_{1/2}(\mu-\nu')\tau(\nu'),
\nonumber\\
\tau(\nu) + \tau_h (\nu) &=&
 \int_{-B'}^{B'}d\mu' K_{1/2}(\nu -\mu')\omega(\mu')
  -\int_{-B''}^{B''} d\nu' K_1 (\nu -\nu')\tau(\nu'),
\label{eq:densityequation}
\end{eqnarray}
%
\begin{multicols}{2}\noindent
where $K_{\rho}(x):=\pi^{-1}\rho/(\rho^2 + x^2)$,  
and $B$, $B'$, and $B''$
in the definite integrals should be 
determined self-consistently.
In the absence of the complex roots, 
$M/N =\int^B_{-B}\sigma(\lambda)d\lambda$,
$M'/N =\int^{B'}_{-B'}\omega(\mu)d\mu$, and
$M''/N =\int^{B''}_{-B''}\tau(\nu)d\nu$.
Once the density $\sigma$ is solved
from eq.(\ref{eq:densityequation}), we have
the $z$-components of the total spin and the total orbital
$$
 \frac{S^z_{tot}}{N}
  =\frac{1}{2} + \int^{B'}_{-B'}\omega(\mu)d\mu 
   -\int^B_{-B}\sigma(\lambda)d\lambda
    -\int^{B''}_{-B''}\tau(\nu)d\nu,
$$
\begin{equation}
\frac{T^z_{tot}}{N}
  =\frac{1}{2}-\int_{-B'}^{B'}\omega(\mu)d\mu,\hspace{36mm}
\label{eq:totalspin}
\end{equation}
the energy 
$$E=-2\pi NJ\int_{-B}^BK_{1/2}(\lambda)\sigma(\lambda)d\lambda,$$
and the momentum
$$ P=-N\int_{-B}^B
        [2\tan^{-1}(2\lambda)-\pi]\sigma(\lambda)d\lambda.$$

\section{The ground state}

The ground state is described
by the densities $\sigma_0(\lambda)$, $\omega_0(\mu)$, 
and $\tau_0(\nu)$ with no
holes and by   $B_0=B'_0=B''_0 \rightarrow \infty$. 
This is true because all the states with holes will have  
higher energies.
In this case, 
eq. (\ref{eq:densityequation}) can be solved. Let
\begin{eqnarray}
\sigma_0(\lambda)=\displaystyle\frac{1}{2\pi}\int_{-\infty}^{\infty}
  \tilde{\sigma}_0(q) e^{-iq\lambda}dq, 
\nonumber\\
\omega_0(\mu)=\displaystyle\frac{1}{2\pi}\int_{-\infty}^{\infty}
  \tilde{\omega}_0(q) e^{-iq\mu}dq,  
\nonumber\\
\tau_0(\nu)=\displaystyle\frac{1}{2\pi}\int_{-\infty}^{\infty}
   \tilde{\tau}_0(q) e^{-iq\nu}dq,
\nonumber
\end{eqnarray}
then
\begin{eqnarray}
	\tilde{\sigma}_0(q) &=& \sinh(3q/2)/\sinh(2q),
	\nonumber\\
	\tilde{\omega}_0(q) &=& \sinh(q)/\sinh(2q),
	\nonumber\\
	\tilde{\tau}_0(q) &=& \sinh(q/2)/\sinh(2q).
	\nonumber
\end{eqnarray}
Hence from eq. (\ref{eq:weight}), 
the highest weight labeling the ground state
is the null vector $(0, 0, 0)$, and so
the ground state is 
a $SU(4)$ singlet.  
This agrees to the theorem of Affleck
and Lieb \cite{AffleckLieb}, a generalization of the spin chain problem
\cite{LiebSchulzMattis}. 
In that state, the total orbital, the total spin,
and their products are
all zero, {\it i.e.},
\begin{eqnarray}
\frac{T^z_{tot}}{N}
   =\frac{1}{2}-\int_{-\infty}^{\infty}\omega_0(\mu)d\mu
    =\frac{1}{2} -\tilde{\omega}_0(0)=0, 
      \nonumber\\
\frac{S^z_{tot}}{N}= \frac{1}{2}+\tilde{\omega}_0(0)-\tilde{\sigma}_0(0)
- \tilde{\tau}_0(0)=0, 
	\nonumber\\
\sum_{j=1}^{N}T_j^zS_j^z =0. \nonumber
\end{eqnarray}
In deriving this result, eqs.(\ref{eq:chevalley}) have been used.
The energy and the momentum
of the ground state are
\begin{eqnarray}
E_0 &=& -NJ(\frac{3}{2}\ln 2 + \frac{\pi}{4})\nonumber\\
P_0 &=& \left\{
           \begin{array}{ll}
             0 \;  {\rm mod }\, 2\pi, & {\rm for}\, N/4={\rm even},\\
             \pi\; {\rm mod }\, 2\pi, & {\rm for}\, N/4={\rm odd}.
           \end{array}
         \right.             
\label{eq:Groundenergy}
\end{eqnarray}
Eq. (\ref{eq:Groundenergy}) coincides with the result of Sutherland
\cite{Sutherland75} after correcting for the trivial overall constant
shift $JN$ between the two models.

\section{Low-lying excitations}

\subsection{Spectra of elementary excitations}

The possible elementary excitation modes are obtained 
by the
variation in the sequence of quantum numbers
$\{ I_j \}$, $\{J_\gamma \}$ or $\{ K_c \}$
from the ground state.
We can assume $B=B'=B'' \rightarrow \infty$ for the low-lying excitations.
The simple modes will be solved by placing
holes in the rapidity-configurations.
If we let $\sigma(\lambda)=\sigma_0(\lambda)+\sigma_1(\lambda)/N$,
$\omega(\mu)=\omega_0(\mu)+\omega_1(\mu)/N$ and
$\tau(\nu)=\tau_0(\nu)+\tau_1(\nu)/N$,
then the excitation energy and momentum
\begin{eqnarray}
   \Delta E = -2\pi J\int_{-\infty}^{\infty}
        K_{1/2}(\lambda)\sigma_1(\lambda)d\lambda,
            \nonumber\\
   \Delta P = -\int_{-\infty}^{\infty}
        [2\tan^{-1}(2\lambda)-\pi]\sigma_1(\lambda)d\lambda,
\end{eqnarray}
and $\Delta M =\int\sigma_1(\lambda)d\lambda$,
$\Delta M'=\int\omega_1(\mu)d\mu$, and
$\Delta M''=\int\tau_1(\nu)d\nu$.
After solving 
the integral equations
(\ref{eq:densityequation}) with
$\sigma_h(\lambda)=\delta(\lambda -\bar{\lambda})/N$,
$\omega_h(\mu)=\delta(\mu -\bar{\mu})/N$ or
$\tau_h(\nu)=\delta(\nu -\bar{\nu})/N$ respectively, 
one finds that
there are
three types of elementary excitation modes.
A hole in the $\lambda$-configuration, $\mu$-configuration,
or $\nu$-configuration ($\lambda$-hole, $\mu$-hole, and 
$\nu$-hole respectively hereafter) creates a $SU(4)$ multiplet labeled
by the highest weight $(1, 0, 0)$, $(0, 1, 0)$ or
$(0, 0, 1)$ respectively. The $\mu$-hole has
a six dimensional representation, the $\lambda$- and $\nu$-holes have four
dimensional representations.  We call these excitations the 
$SU(4)$ flavorons.
The energies of these elementary excitation are
\def\cosh{\,{\rm cosh }\,}
\def\sinh{\,{\rm sinh }\,}
\def\tanh{\,{\rm tanh }\,}
\def\tan{\,{\rm tan }\,}%
\begin{eqnarray}
\varepsilon_\sigma(\bar{\lambda})
 &=&\frac{J\pi/2}{\sqrt{2}\cosh (\bar{\lambda}\pi/2)-1},
   \nonumber\\
\varepsilon_\omega (\bar{\mu}) 
  &=& \frac{J\pi/2}{\cosh (\bar{\mu}\pi/2)},
  \nonumber\\
\varepsilon_\tau (\bar{\nu})
  &=& \frac{J\pi/2}{\sqrt{2}\cosh(\bar{\nu}\pi/2)+1},
\label{eq:excitenergy}
\end{eqnarray} 
where $\bar{\lambda}$, $\bar{\mu}$ and $\bar{\nu}$ stand
for the positions of holes in the corresponding rapidity 
configurations. These excitation energies vanish when the
positions of holes go to infinity in
the thermodynamic limit. Therefore they are
gapless modes. The momenta of the excitations are given by

\begin{eqnarray}
p_\sigma (\bar{\lambda}) &=&
   2\tan^{-1}\left[(\sqrt{2}+1)\tanh(\bar{\lambda}\pi/4)\right]-3\pi/4,
               \nonumber\\
p_\omega (\bar{\mu}) &=&
   2\tan^{-1}\left[ \tanh (\bar{\mu}\pi/4) \right]-\pi/2,
               \nonumber\\
p_\tau (\bar{\nu}) &=&
  2\tan^{-1}\left[(\sqrt{2}-1)\tanh(\bar{\nu}\pi/4)\right]-\pi/4.
\label{eq:momentum}
\end{eqnarray}
Eliminating the rapidities in eqs. (\ref{eq:excitenergy}) 
and (\ref{eq:momentum}), we have
\begin{eqnarray}
\varepsilon_\sigma (p_\sigma)&=&
   \frac{J\pi}{2}\left[\sqrt{2}\cos(p_\sigma+3\pi/4) +1\right],
             \nonumber\\
\varepsilon_\omega (p_\omega) &=&
   \frac{J\pi}{2}\cos(p_\omega+\pi/2),
              \nonumber\\
\varepsilon_\tau (p_\tau)&=&
   \frac{J\pi}{2}\left[\sqrt{2}\cos(p_\tau+\pi/4) -1\right],
\label{eq:dispersion}
\end{eqnarray}
where $p_\sigma\in [-3\pi/2,0],$
$p_\omega \in [-\pi,0],$
$p_\tau \in [-\pi/2,0].$ 

\begin{figure}
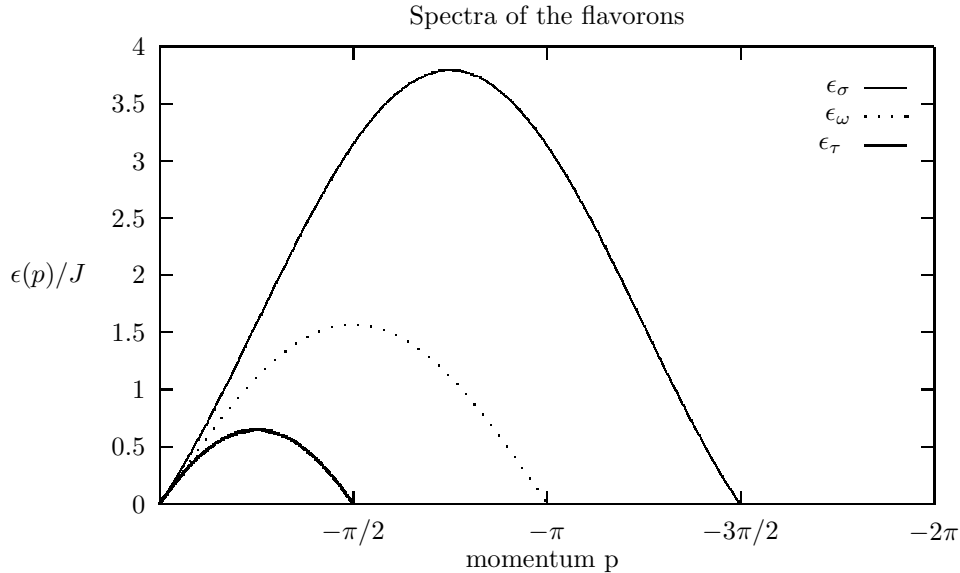

\setlength{\unitlength}{1cm}

\caption{
The spectra of the three type quasi-particles}
\label{fig:fspectra}
\end{figure}

We are now in the position to relate the elementary
excitations to the spin and
orbital in the original model, eq. (\ref{eq:Hamiltonian}).
The quadruplets $(1, 0, 0)$ or $(0, 0, 1)$ are
flavorons carrying both spin $1/2$ and orbital $1/2$
with energies $\epsilon_\sigma$ or $\epsilon_\tau$,
the  hexaplet $(0, 1, 0)$  describes flavoron
s
carrying either spin $1$ or orbital $1$ with energy $\epsilon_\omega$.
The spectra of these three types of excitations are plotted in 
fig.(\ref{fig:fspectra}). 
In comparison to Sutherland's results,
eq.(\ref{eq:dispersion}) differs only in that
each mode is shifted by a different 
constant in momentum arising from the
$\pi$ term in Eq. (\ref{eq:momentumf}). 
It seems more convenient to use
the present version,  eq. (\ref{eq:dispersion}), 
to study the spectra of the
magnon-type
excitations
in the subsection V.C.
\subsection{The complex roots} 

Because of the existance of the complex roots \cite{Woynarovich}
in the solution set of the Bethe ansatz equations (\ref{eq:BAE}),
we must consider their contributions, particularly from the 2-strings.
In this case we need to rederive the integral 
equations (\ref{eq:densityequation}). We obtain the same
equation formally but now the inhomogeneous terms $\sigma_h(\lambda)$,
$\omega_h(\mu)$, and $\tau_h(\nu)$ include also the contributions from the
complex roots. A 2-string in the $\lambda$-configuration,
$\lambda_{\pm} = \lambda_0 \pm i/2$, introduces additional terms 
in eq. (\ref{eq:densityequation}). As a result, we
have
\begin{eqnarray} 
\sigma_h(\lambda) 
   &=& [K_{3/2}(\lambda-\lambda_0)+K_{1/2}(\lambda-\lambda_0)]/N, 
                         \nonumber\\ 
\omega_h(\mu) &=& -K_1(\mu-\lambda_0)/N, \nonumber\\ 
\tau_h(\nu) &=& 0. 
\end{eqnarray} 
The energy is given by
\begin{eqnarray} 
\Delta E &=& -2\pi J\int_{-\infty}^{\infty}  
             K_{1/2}(\lambda)\sigma_1(\lambda)d\lambda  
                         \nonumber\\ 
    \,  &\,& -2\pi J[K_{1/2}(\lambda_0+i/2)+K_{1/2}(\lambda_0-i/2)], 
\label{eq:stringenergy} 
\end{eqnarray} 
and the integral equation leads to 
\begin{eqnarray} 
	\tilde{\sigma_1}(q)=-\exp(iq\lambda_0-|q|/2). 
	\nonumber 
\end{eqnarray} 
Our calculation shows a complete cancellation in 
eq.(\ref{eq:stringenergy}). Therefore a 2-string 
in the $\lambda$-configuration does not change the energy. 
We also find $M=N\int\sigma(\lambda)d\lambda+2$, 
where $\sigma(\lambda)$ is the
density of the real roots. 

Similar results are found for the 2-strings in the $\nu$- and
$\mu$-configurations.  The equations for a $\nu$-configuration are given by, 
\begin{eqnarray} 
\sigma_h(\lambda)&=&0,\nonumber\\ 
\omega_h(\mu)&=&-K_1(\mu-\nu_0)/N,\nonumber\\ 
\tau_h(\nu)&=&[K_{3/2}(\nu-\nu_0)+K_{1/2}(\nu-\nu_0)]/N \nonumber 
\end{eqnarray} 
and  
$M''=N\int\tau(\nu)d\nu+2$. 
By solving  $\sigma_1 (\lambda)=0$, we obtain 
\begin{eqnarray} 
	\Delta E = -2\pi J\int K_{1/2}\sigma_1(\lambda)d\lambda = 0. 
	\nonumber 
\end{eqnarray} 
And the equations for a 2-string in the $\mu$-configuration are given by,  
\begin{eqnarray} 
\sigma_h(\lambda)&=&-K_1(\lambda-\mu_0)/N, \nonumber\\ 
\omega_h(\mu)&=&[K_{1/2}(\mu-\mu_0)+K_{3/2}(\mu-\mu_0)]/N,\nonumber\\  
\tau_h(\nu)&=&-K_1(\nu-\mu_0)/N, \nonumber  
\end{eqnarray} 
and  
$M'=N\int\omega(\mu)d\mu +2$, 
we obtain $\sigma_1(\lambda)=0$ and hence $\Delta E=0$.  
Although these three types of the 2-strings do not contribute to the
energy, they do contribute to the quantum numbers of spin and orbital, 
and to the highest weight of the $SU(4)$ representations.

\subsection{Generalized magnon type excitations}
\label{sec:excitations}

The flavorons discussed in the previous subsection
are elementary excitations
of the system.  These flavorons may combine to form composite
excitations similar to the magnon excitations in the 1-dimensional spin
chain\cite{Mattis},
which are of interest to experiments and numerical simulations.
In such a construction, the  structure of the decomposition of the direct
product of the $SU(4)$ fundamental representation must be taken into account.
Let us  consider $N=4n$, the decomposition brings about  a direct sum of a
series of irreducible representations, i.e., $(0,0,0)$, $(1,0,1)$, $(0,2,0)$,
$(2,1,0)$, $(4,0,0)$ $\it{etc.}$  The composite excitation states include both
the singlet $(0,0,0)$  and the multiplets of 15-fold $(1,0,1)$, of 20-fold
$(0,2,0)$, and of 45-fold $(2,1,0)$ or of 35-fold $(4,0,0)$ etc.  Those
multifold excitations are the generalization of magnon excitations to the spin
systems with orbital degeneracy.  One $\lambda$-hole and one
$\nu$-hole together create a 15-fold multiplet with excitation energy and
momentum, 
\begin{eqnarray}
\Delta E_{(15)}= \varepsilon_\sigma (\bar{\lambda}) +
\varepsilon_\tau (\bar{\nu}), 
\nonumber\\ 
\Delta
P_{(15)}=  p_\sigma (\bar{\lambda})+
           p_\tau (\bar{\nu}), 
\nonumber
\end{eqnarray} 
which is a pair of flavorons of $\sigma$-type and $\tau$-type.
Two $\mu$-holes create a 20-fold multiplet with
\begin{eqnarray}
\Delta E_{(20)}=
     \varepsilon_\omega (\bar{\mu}_1)
          +\varepsilon_\omega (\bar{\mu}_2),
\nonumber\\ 
\Delta P_{(20)}=
   p_\omega (\bar{\mu}_1)
   + p_\omega (\bar{\mu}_2).
\nonumber
\end{eqnarray} 
The 45-fold multiplet is a three-hole state created by 
two $\lambda$-holes and one $\mu$-hole, for which the excitation
energy and momentum are  
\begin{eqnarray}
\Delta E_{(45)}=
   \varepsilon_\sigma(\bar{\lambda}_1)
     +\varepsilon_\sigma(\bar{\lambda}_2)
       +\varepsilon_\omega(\bar{\mu}),
\nonumber\\ 
\Delta P_{(45)}=
  p_\sigma(\bar{\lambda}_1)
     +p_\sigma(\bar{\lambda}_2)
       +p_\omega(\bar{\mu}).
\nonumber
\end{eqnarray} 
Four $\lambda$-holes create a 35-fold multiplet with
\begin{eqnarray}
\Delta E_{(35)}=\sum_{j=1}^4
     \varepsilon_\sigma(\bar{\lambda}_j),
\nonumber\\ 
\Delta P_{(35)}=\sum_{j=1}^4
     p_\sigma(\bar{\lambda}_j).
\nonumber
\end{eqnarray} 
The singlet excitation is obtained by placing
a $\lambda$-hole, a $\nu$-hole and three 2-strings in 
$\lambda$, $\mu$ and $\nu$-configurations respectively.
The singlet is degenerate with the 15-fold multiplet in 
energy, i.e. 
\begin{eqnarray}
\Delta E_{(1)}= \varepsilon_\sigma(\bar{\lambda})
 +\varepsilon_\tau(\bar{\nu}).
\nonumber
\end{eqnarray} 
In the above equations,  $\epsilon_\sigma(\bar{\lambda})$, 
$\epsilon_\omega(\bar{\mu})$ and $\epsilon_\tau(\bar{\nu})$
are given by eq.(\ref{eq:excitenergy}),
and $p_\sigma(\bar{\lambda})$, 
$p_\omega(\bar{\mu})$, and $p_\tau(\bar{\nu})$
are given by eq.(\ref{eq:momentum}).
  
\begin{figure}
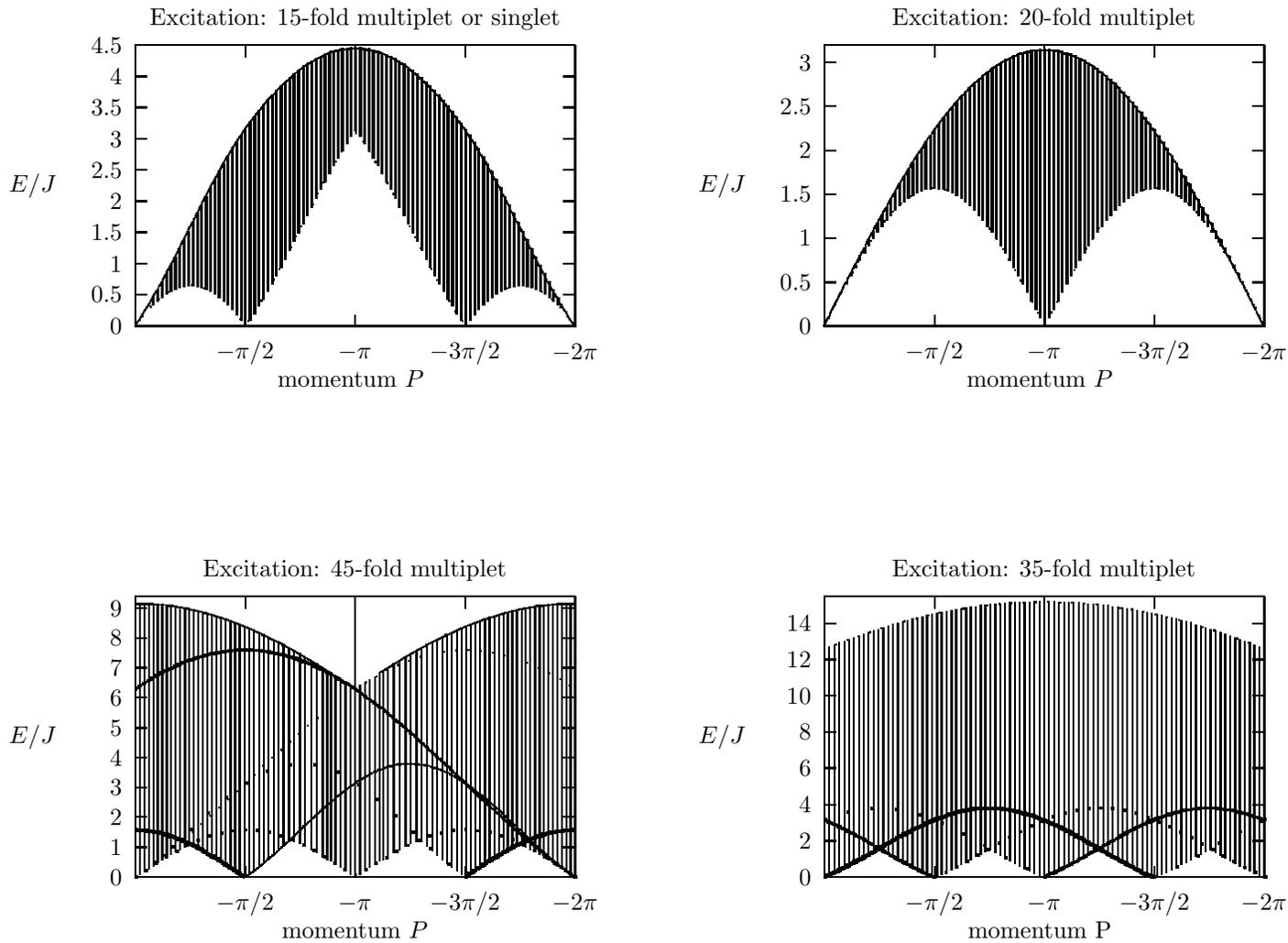

\setlength{\unitlength}{1cm}
%
\caption{
The dispersions for various types of the generalized magnons.
$E$ and $P$ stand for the the corresponding 
$\Delta E$ and $\Delta P$ in Sec.\ref{sec:excitations}}
\label{fig:dispersion}
\end{figure}

The energy-momentum dispersion of various magnon-types
of excitations are plotted in Fig. (\ref{fig:dispersion}). 
In the spectrum
calculations, we have used the 
periodicity in momentum $P$, so that 
$\Delta E (P +2\pi) = \Delta E (P)$.  
For instance, for the 45-fold 
degenerate states, 
$\Delta E_{(45)} (P) =
\epsilon_{\sigma}(q_1) + 
\epsilon_{\sigma}(q_2) +
\epsilon_{\omega}(q_3)$, where
$P=q_1+q_2+q_3$, with modula $2\pi$. 
In a recent paper in Ref. 15, 
the lower-lying excitations of model (\ref{eq:Hamiltonian}) 
were calculated 
numerically for finite
systems. Their results 
are consistent with ours 
in Fig. 2. In particular, both the numerical 
calculations and
the present Bethe ansatz solutions show 
the following feature: as the momentum
$|p|$ increases from $0$ to $\pi$, 
the lowest excitations are changed
 from the 15-fold degenerate states to the
45-fold degenerate states 
at $|P|=\pi/2$. 

\begin{figure}
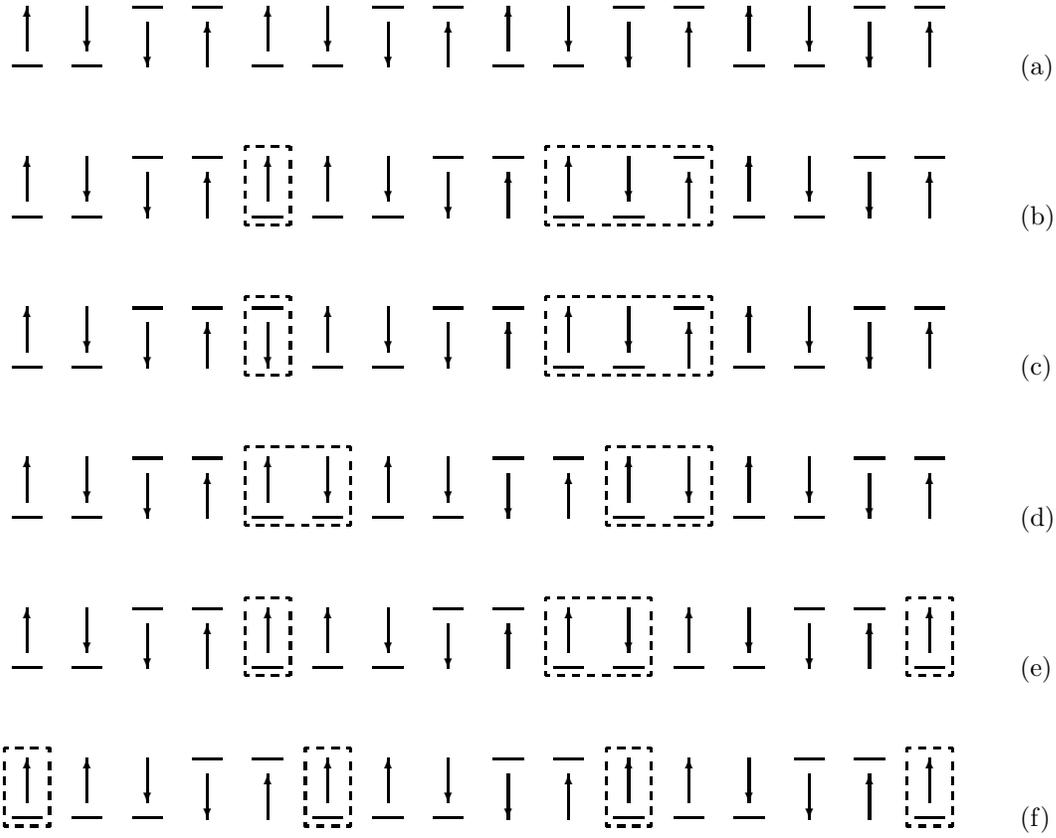

%
\vspace{10mm}
\caption{
(a). A $<S_j T_j>$ staggerily ordered state to demonstrate the
ground state, the SU(4) singlet.
(b). flavorons of one $\sigma$-type and one $\tau$-type compounded
symmetrically to form a 15-fold multiplet, and (c). compounded
anti-symmetrically to form a singlet.
(d). The 20-fold multiplet being compounded of two flavorons
of $\omega$-type.
(e). The 45-fold multiplet being compounded of two flavorons 
of $\sigma$-type and one flavorons of $\omega$-type.
(f). The 35-fold multiplet being compounded of four flavorons
of $\sigma$-type.
}
\label{fig:flavoron}
\end{figure}

In Fig. \ref{fig:flavoron}, we illustrate  
these generalized magnon types 
of composite low lying
excited states.  We start with a typical configuration of the 
the ground state in Fig. \ref{fig:flavoron} (a), 
and the various generalized magnon excitations
are created from the ground state by introducing two or more 
flavorons as shown in 
(b)-(f), which arise from various possible flips of spin or orbital or
both.
 The flavoron indicated by the 
dashed box in fig.(\ref{fig:flavoron}) moves in the background
of the $SU(4)$ singlet carrying both energy and the quantum numbers.
The $\sigma$-type (flavoron) excitation mode is a moving quaduplet
with $|\tup>$ being the local highest weight state. 
The $\omega$-type excitation mode is a moving hexaplet
with $|\tup\tdown> - \,|\tdown\tup>$ being the 
local highest weight state. 
The $\tau$-type excitation mode is a moving quaduplet with
$|\tup\tdown\bup> - \,|\tup\bup\tdown>
+ \,|\tdown\bup\tup> - \,|\tdown\tup\bup>
+ \,|\bup\tup\tdown> - \,|\bup\tdown\tup>$
being the local highest weight state.

\section{Summary and Acknowledgment}
In this paper we have used Bethe ansatz method to discuss extensively the
ground state
and various types of the low lying excited states of a Heisenberg spin chain
with two-fold orbital degeneracy in the limit of $SU(4)$ symmetry. There
are three types of elementary 
excitations in the present model.  
Two of them
carry spin 1/2 and orbital 1/2,
and both are four-fold degenerate. The third
one carries either spin 1 or orbital 1, and 
is six-fold degenerate. We have constructed magnon
types of composite  excitations and calculated their spectra.  

We thank G.S. Tian for a reference.
YQL acknowledges grants NSFC-19675030, NSFZ-198024 and the support from  
the Y. Pao \& Z. Pao Foundation.  The work is supported in part by DOE 
grant No. DE/FG03-98ER45687.

\end{multicols}

\newpage

\newpage

\unitlength 2mm
\linethickness{0.8pt}

\end{document}